\documentstyle[preprint,aps,eqsecnum]{revtex}
 
\def\beq#1{\begin{equation} \label{#1}}
\def\eeq{\end{equation}}

\def\ket#1{\left\vert #1\right\rangle}

\begin{document}
{
\tighten
\preprint {\vbox{
\hbox{WIS-98/26/Oct-PH}
\hbox{TAUP 2527-98}
\hbox{ANL-HEP-PR-98-123}
\hbox{hep-ph/9809347}
}}
 
\title{A useful approximate isospin equality for charmless strange B Decays }

\author{Harry J. Lipkin\,\thanks{Supported
in part by The German-Israeli Foundation for Scientific Research and 
Development (GIF) and by the U.S. Department
of Energy, Division of High Energy Physics, Contract W-31-109-ENG-38.}}
\address{ \vbox{\vskip 0.truecm}
  Department of Particle Physics
  Weizmann Institute of Science, Rehovot 76100, Israel \\
\vbox{\vskip 0.truecm}
School of Physics and Astronomy,
Raymond and Beverly Sackler Faculty of Exact Sciences,
Tel Aviv University, Tel Aviv, Israel  \\
\vbox{\vskip 0.truecm}
High Energy Physics Division, Argonne National Laboratory,
Argonne, IL 60439-4815, USA\\
~\\HJL@axp1.hep.anl.gov
\\~\\
}
 
\maketitle
  
\begin{abstract}

A useful inequality is obtained if charmless strange B decays are assumed to be
dominated by a $\Delta I = 0$ transition like that from the gluonic penguin 
diagram and the contributions of all other diagrams including the tree, 
electroweak penguin and annihilation diagrams are small but not negligible. 
The interference contributions which are linear in these other amplitudes are 
included but the direct contributions which are quadratic are neglected. 

\end{abstract}
} 

It is now believed that charmless strange B decays are dominated by the gluonic
penguin diagram. If all other diagrams are negligible, the branching ratios for
sets of decays to states in the same isospin multiplets are uniquely related by
isospin because the gluonic penguin leads to a pure I=1/2 final state, and there
is no simple mechanism in the standard model that can give CP violation.

If the tree diagram produced by the $\bar b \rightarrow \bar u u \bar s$ 
transition at the quark level also contributes, the isospin analysis becomes
non-trivial, much more interesting and CP violation can be observed. This case
has been considered in great detail by Nir and Quinn\cite{PBPENG}.

Our purpose here is to point out and discuss an intermediate case, where the 
amplitude from the tree diagram is not negligible but still sufficiently small 
by comparison with the gluonic penguin that its contribution to branching ratios
need be considered only to first order, and higher order contributions can be 
neglected. An interesting approximate sum rule is obtained which can check the
validity of this approximation and guide the search for CP violation. 

We consider here the  $ B \rightarrow K \pi $ decays. Exactly the same 
considerations hold for all similar decays into a strange meson and an isovector
nonstrange meson. An extension incorporating also the decay into the isoscalar
strange  meson; e.g. $\omega$ is presented below for all ideally mixed nonets.

The gluonic penguin diagram leads to a pure isospin 1/2 state. Its contributions
to all $ B \rightarrow K \pi $ amplitudes, denoted by $P$ are simply related by 
isospin. 
$$ P(B^+ \rightarrow K^+ \pi^o) =
- P (B^o \rightarrow K^o \pi^o) = 
{1 \over \sqrt 2} \cdot P (B^o \rightarrow K^+ \pi^-) =  
{1 \over \sqrt 2} \cdot P (B^+ \rightarrow K^o \pi^+) \equiv P
\eqno(1)                                          $$
The tree diagram has two independent contributions corresponding to the 
color-favored and color-suppressed couplings of the final quark-antiquark pairs.
Their amplitudes, denoted by $T_f$ and $T_s$ are also simply related by isospin.
$$ {1 \over \sqrt 2} \cdot T_f (B^o \rightarrow K^+ \pi^-) = 
T_f (B^+ \rightarrow K^+ \pi^o) 
\equiv T_f 
\eqno(2a)                      $$
$$ T_s (B^+ \rightarrow K^+ \pi^o) =
T_s (B^o \rightarrow K^o \pi^o)  
\equiv T_s 
\eqno(2b)                  $$
In the approximation where we consider the contributions of the tree amplitudes 
only to first order, we obtain:
$$ BR (B^o \rightarrow K^+ \pi^-) \approx 2 P^2 + 4 P \cdot T_f
\eqno(3a)                                          $$
$$ BR (B^o \rightarrow K^o \pi^o) \approx P^2 - 2  P \cdot T_s
\eqno(3b)                                          $$
$$ BR (B^+ \rightarrow K^o \pi^+) \approx 2 P^2 
\eqno(3c)                                          $$
$$ BR (B^+ \rightarrow K^+ \pi^o) \approx P^2 + 2 P \cdot T_f
+ 2 P \cdot T_s 
\eqno(3d)                                          $$
This leads to the approximate equality 
$$ BR (B^o \rightarrow K^+ \pi^-) -2 BR 
(B^o \rightarrow K^o \pi^o) \approx 
2 BR (B^+ \rightarrow K^+ \pi^o) - BR (B^+ \rightarrow K^o \pi^+) \approx $$
$$ \approx 4 P \cdot T_f + 4 P \cdot T_s = 
4 P(T_f + T_s)\cdot \rm cos (\phi_P - \phi_T - \phi_S)
\eqno(4a)                                          $$
where $\phi_P$ and $\phi_T$ are the weak phases respectively of the penguin and
tree amplitudes and $\phi_S$ is the strong phase difference between the two.
Note that both the left-hand and right hand sides vanish independently for
any transition like the pure gluonic penguin that leads to a pure I=1/2 state.
The relation (4) therefore is due entirely to interference between the 
penguin I=1/2 amplitude and the I=3/2 component of the tree amplitude.
This immediately leads to the correspnding expression for the charge-conjugate
decays,
$$ BR (\bar B^o \rightarrow K^- \pi^+) -2 BR 
(\bar B^o \rightarrow \bar K^o \pi^o) \approx 
2 BR (B^- \rightarrow K^- \pi^o) - BR (B^- \rightarrow \bar K^o \pi^-) \approx 
$$
$$ 
4 P(T_f + T_s)\cdot \rm cos (\phi_P - \phi_T + \phi_S)
\eqno(4b)                                          $$
The direct CP violation is seen to be given by
$${{2 BR(B^- \rightarrow K^- \pi^o) - 2 BR (B^+ \rightarrow K^+ \pi^o) -
BR (B^- \rightarrow \bar K^o \pi^-) + BR (B^+ \rightarrow K^o \pi^+)} \over
{2 BR(B^- \rightarrow K^- \pi^o) + 2 BR (B^+ \rightarrow K^+ \pi^o) -
BR (B^- \rightarrow \bar K^o \pi^-) - BR (B^+ \rightarrow K^o \pi^+)}} 
\approx $$
$$\approx 
 \rm tan (\phi_P - \phi_T) \rm tan ( \phi_S)
\eqno(5)                                          $$
The same equality (4) is easily seen in the formalism used by Nir and 
Quinn\cite{PBPENG} using their three amplitudes denoted by $U$, $V$ and $W$.
The $W$ amplitude is the penguin amplitude and the the two tree amplitudes
$U$ and $V$ are linear combinations of our $T_f$ and $T_s$ amplitudes, 
defined by isospin properties rather than quark diagrams.  
$$ BR (B^o \rightarrow K^+ \pi^-) -2 BR (B^o \rightarrow K^o \pi^o) = 
-2W\cdot(U+V) + 2(V^2 - U^2) \approx -2W\cdot(U+V)
\eqno(6a)                                          $$
$$ 2 BR (B^+ \rightarrow K^+ \pi^o) - BR (B^+ \rightarrow K^o \pi^+)
= -2W\cdot(U+V) -  2(V^2 - U^2) \approx -2W\cdot(U+V)
\eqno(6b)                                          $$
It is interesting to note that here also one sees the apparent miracle that
the $same$ linear combination of the two tree amplitudes, $U+V$ appears in 
both expressions (6a) and (6b) and that the approximate equality follows
from the approximation that quadratic terms in U and V are negligible in
comparison with the product of the linear terms and the dominant penguin
amplitude W.  
$$ (V^2 - U^2) <<  W\cdot(U+V)    \eqno(7)               $$

However, a simple isospin analysis shows that this is no miracle and that the 
approximate equality (4) holds also when contributions of
annihilation diagrams, diagrams in which a flavor-changing final state 
interaction like charge exchange follows the weak tree diagram and electroweak 
penguins. We first note that both the annihilation diagram and the
charge exchange diagram which proceeds via the quark annihilation and pair 
creation transition $u \bar u \rightarrow $ gluons $\rightarrow d \bar d$ lead
to pure I=1/2 final states and their contributions cancel in the sum rule (4).

The electroweak penguin amplitudes contain both I=1/2 and I=3/2 components.
This amplitude which we denote by $P_{EW}$ can be written as the sum of a 
contribution in which the electroweak boson creates an isoscalar $q \bar q$ 
pair and one which we denote by $P_u$ in which the electroweak boson creates a 
$u \bar u$ pair. The contribution with the isoscalar pair is proportional to 
the gluonic penguin.
Thus 
$$ P_{EW} = \xi \cdot P + P_u                          \eqno (8)  $$
where $\xi$ is a small parameter. To first order in $P_{EW}$ the contribution of
the $\xi \cdot P$ term satisfies the approximate equality (4), since the 
contributions of both sides vanish. The $P_u$ term arises from the quark 
transition 
$$ \bar b \rightarrow \bar s u \bar u \eqno(9a) $$
This then leads to the hadronic transition
$$ B(\bar b q) \rightarrow \bar s u \bar u q \eqno(9b) $$
But these are exactly the same as the tree transitions (2) except that the 
color favored and color suppressed transitions are reversed. Here
the $u \bar u$ is a color singlet and the transitions in which this $u \bar u$
pair combine to make a $\pi^o$ are color favored. 
We can therefore write $P_u$ as the sum of a color-favored and color suppressed
term,
$$ P_u = P_{uf} + P_{us} \eqno(10)     $$
and note that they satisfy the isospin relations analogous to (2),
$$ {1 \over \sqrt 2} \cdot P_{us} (B^o \rightarrow K^+ \pi^-) = 
P_{us} (B^+ \rightarrow K^+ \pi^o) 
\equiv P_{us} 
\eqno(11a)                      $$
$$ P_{uf} (B^+ \rightarrow K^+ \pi^o) =
P_{uf} (B^o \rightarrow K^o \pi^o)  
\equiv P_{uf} 
\eqno(11b)                  $$
We can therefore generalize the equalities (4) by replacing $T_f$ and $T_s$
by
$$ T_f \rightarrow T'_f \equiv T_f + P_{us} \eqno(12a)  $$
$$ T_s \rightarrow T'_s \equiv T_f + P_{uf} \eqno(12b)  $$
We thus see that the inclusion of the electroweak penguin has the effect of 
adding a term with the same isospin properties as the gluonic penguin and terms
with the isospin properties of the color-suppressed and color-favored tree 
diagrams, and therefore do not affect the approximate equality (4).

The reason for the equality is easily seen by noting the isospin properties of
all charmless strange decays. The weak interaction produces either a 
$\Delta I=0$ transition which leads to an $I=1/2$ final state or a $\Delta I=1$
transition which can lead to both $I=1/2$ and $I=3/2$ final states\cite{GHLRP}.
There are therefore three independent isospin amplitudes. The four decay 
branching ratios therefore depend in the general case on five parameters, three
magnitudes and two relative phases and no simple equality between the branching 
ratios is obtainable. However, in the approximation where only linear terms in 
the $\Delta I = 1$ amplitudes are considered, a phase convention can be chosen 
in which the dominant $\Delta I = 0$ amplitude is real and only the real parts 
of the $\Delta I = 1$ amplitudes contribute. The four decay branching ratios
now depend on only three real parameters, and the equality (4) is therefore
obtained.

The same approach can be applied to charmless decays to a strange pseudoscalar
and a nonstrange vector meson. Here the ideal mixing of the $\rho$ and $\omega$ 
enables inclusion of the $K\omega$ final states in the analysis if the OZI rule
is assumed\cite{bkpfsi}.
We  first express the $\rho^o$ and $\omega$ states in terms of their $u \bar u$
and $d \bar d$ components, denoted respectively as $V_u$ amd $V_d$,
$$ \ket{\rho^o} \equiv {1 \over \sqrt 2} \cdot (\ket{V_u} - \ket{V_d});
~ ~ ~ \ket{\omega} \equiv {1 \over \sqrt 2} \cdot (\ket{V_u} + \ket{V_d}) 
\eqno(13) $$
The analogs of eqs. (1-2) have a particularly simple form
when written in terms of the flavor eigenstates $V_u$ amd $V_d$,
$$ P(B^+ \rightarrow K^+ V_u) =P(B^o \rightarrow K^o V_d) =
P (B^o \rightarrow K^+ \rho^-) =  P (B^+ \rightarrow K^o \rho^+) 
\equiv {\sqrt 2} \cdot P
\eqno(14a)                                          $$
$$ T_f (B^o \rightarrow K^+ \rho^-) = 
T_f (B^+ \rightarrow K^+ V_u) 
\equiv {\sqrt 2} \cdot T_f 
\eqno(14b)                      $$
$$ T_s (B^+ \rightarrow K^+ V_u) 
= T_s (B^o \rightarrow K^o V_u) \equiv T_s 
\eqno(14c)                  $$
The negative signs and $\sqrt 2$ factors which appear in eqs.
(1-2) as isospin Clebsch-Gordan coefficients are completely absent in this
quark-flavor representation. These factors appear in the quark model description
only in the wave functions of the $\rho^o$ and $\pi^o$. This description 
enables extending the treatment beyond isospin and also including the $\omega$.
This implicitly asssumes the OZI rule which discards the penguin diagram in 
which three gluons create the omega\cite{bkpfsi}. It also neglects the similar 
electroweak diagrams in which the electroweak boson creates the $\omega$. 
Substituting the $\omega$ and $\rho$ wave functions from eq. (13) into the 
relations (14) brings them to the form of eqs. (1-2) with the additional 
predictions for the $K \omega$ decays. 

$$ P(B^+ \rightarrow K^+ \rho^o) = P(B^+ \rightarrow K^+ \omega) =
- P(B^o \rightarrow K^o \rho^o) = P(B^o \rightarrow K^o \omega) = $$
$$ = {1 \over \sqrt 2} \cdot P (B^o \rightarrow K^+ \rho^-) =  
{1 \over \sqrt 2} \cdot P (B^+ \rightarrow K^o \rho^+) \equiv P
\eqno(15a)                                          $$
$$ {1 \over \sqrt 2} \cdot T_f (B^o \rightarrow K^+ \rho^-) = 
T_f (B^+ \rightarrow K^+ \rho^o) = T_f (B^+ \rightarrow K^+ \omega) 
\equiv T_f 
\eqno(15b)                      $$
$$ T_s (B^+ \rightarrow K^+ \rho^o) = T_s (B^+ \rightarrow K^+ \omega) =
T_s (B^o \rightarrow K^o \rho^o) = T_s (B^o \rightarrow K^o \omega)
\equiv T_s 
\eqno(15c)                  $$
In the approximation where we consider the contributions of amplitudes other 
than gluonic penguin only to first order, we obtain:
$$ BR (B^o \rightarrow K^+ \rho^-) \approx 2 P^2 + 4 P \cdot T_f
\eqno(16a)                                          $$
$$ BR (B^o \rightarrow K^o \rho^o) \approx P^2 - 2  P \cdot T_s
\eqno(16b)                                          $$
$$ BR (B^+ \rightarrow K^o \rho^+) \approx 2 P^2 
\eqno(16c)                                          $$
$$ BR (B^+ \rightarrow K^+ \rho^o) = BR (B^+ \rightarrow K^+ \omega)
\approx P^2 + 2 P \cdot T_f + 2 P \cdot T_s 
\eqno(16d)                                          $$
$$ BR (B^o \rightarrow K^o \omega) \approx P^2 + 2  P \cdot T_s
\eqno(16e)                                          $$
This leads again to an approximate equalities 
$$ BR (B^o \rightarrow K^+ \rho^-) -2 BR 
(B^o \rightarrow K^o \rho^o) \approx 
2 BR (B^+ \rightarrow K^+ \rho^o) - BR (B^+ \rightarrow K^o \rho^+) \approx $$
$$ \approx 4 P \cdot T_f + 4 P \cdot T_s = 
4 P(T_f + T_s)\cdot \rm cos (\phi_P - \phi_T - \phi_S)
\eqno(17a)                                          $$
$$ BR (\bar B^o \rightarrow K^- \rho^+) -2 BR 
(\bar B^o \rightarrow \bar K^o \rho^o) \approx 
2 BR (B^- \rightarrow K^- \rho^o) - BR (B^- \rightarrow \bar K^o \rho^-) 
\approx 
$$
$$ 
4 P(T_f + T_s)\cdot \rm cos (\phi_P - \phi_T + \phi_S)
\eqno(17b) $$
where $\phi_P$ and $\phi_T$ are again weak phases respectively of penguin and
tree amplitudes and $\phi_S$ is the strong phase difference between the two.
The relation (17) is again due entirely to interference between the 
penguin I=1/2 amplitude and the I=3/2 component of other amplitudes.

The inclusion of the $\omega$ gives the well-known equality\cite{bkpfsi} 
(16d) and a new approximate equality,
$$ BR (B^o \rightarrow K^o \rho^o) + BR (B^o \rightarrow K^o \omega) \approx 
BR (B^+ \rightarrow K^o \rho^+) \approx 2 P^2 
\eqno(18)                                          $$

\acknowledgments
It is a pleasure to thank Yosef Nir and Jonathan L. Rosner for helpful 
discussions and comments. 
 \def \prd#1#2#3{Phys. Rev. D {\bf#1}, (#3) #2 }
{
\tighten

}
 
\end{document}